\documentclass[twocolumn]{bmcart}

\usepackage[utf8]{inputenc} 
\usepackage{eurosym}
\usepackage{amssymb}
\usepackage{amsmath}
\usepackage{caption}
\usepackage[pdftex]{rotating}
\usepackage{portland}
\usepackage{lastpage}
\usepackage{lipsum}
\usepackage{lipsum}
\usepackage{graphicx}

\startlocaldefs
\endlocaldefs

\begin{document}

\begin{frontmatter}

\begin{fmbox}
\dochead{Research}


\title{Adding new experimental arms to randomised clinical trials: impact on error rates}


\author[
   addressref={aff1},
   corref={aff1},                       
   email={b.choodari-oskooei@ucl.ac.uk}   
]{\inits{BCO}\fnm{Babak} \snm{Choodari-Oskooei}}

\author[
addressref={aff2},
email={danielbratton22@gmail.com}
]{\inits{DJB}\fnm{Daniel J} \snm{Bratton}}

\author[
addressref={aff3},
   email={Melissa.Gannon@lshtm.ac.uk}   
]{\inits{MRS}\fnm{Melissa R} \snm{Gannon}}

\author[
addressref={aff1},
   email={a.meade@ucl.ac.uk}   
]{\inits{Ms}\fnm{Angela M} \snm{Meade}}

\author[
addressref={aff1},
   email={m.sydes@ucl.ac.uk}   
]{\inits{MRS}\fnm{Matthew R} \snm{Sydes}}

\author[
addressref={aff1},
   email={m.parmar@ucl.ac.uk}   
]{\inits{MKB}\fnm{Mahesh KB} \snm{Parmar}}


\address[id=aff1]{
  \orgname{MRC Clinical Trials Unit at UCL, Insttitute of Clinical Trials and Methodology}, 
  \street{90 High Holborn}, 
  \postcode{WC1V 6LJ},                                
  \city{London},                              
  \cny{UK}                                    
}
\address[id=aff2]{%
  \orgname{Clinical Statistics, GlaxoSmithKline, Stockley Park},
  \city{Middlesex},
  \cny{UK}
}
\address[id=aff3]{%
  \orgname{Department of Health Services Research and Policy, LSHTM},
  \city{London},
  \cny{UK}
}


\begin{abstractbox}

\begin{abstract} 
\parttitle{Background} 
Experimental treatments pass through various stages of
development. If a treatment passes through early phase experiments, the
investigators may want to assess it in a late phase randomised controlled
trial. An efficient way to do this is adding it as a new research arm to an
ongoing trial. This allows to add the new treatment while the existing arms 
continue. The familywise type I error rate (FWER) is often a key quantity of 
interest in any multi-arm trial. We set out to clarify how it should be 
calculated when new arms are added to a trial some time after it has started.

\parttitle{Methods} 
 We show how the FWER, any-pair and all-pairs powers can be
calculated when a new arm is added to a platform trial. We extend the
Dunnett probability and derive analytical formulae for the correlation
between the test statistics of the existing pairwise comparison and that of
the newly added arm. We also verify our analytical derivation via
simulations.

\parttitle{Results} 
Our results indicate that the FWER depends on the shared
control arm information (i.e. individuals in continuous
and binary outcomes and primary outcome events in
time-to-event outcomes) from the common control arm patients and the
allocation ratio. The FWER is driven more by the number of pairwise comparisons 
and the corresponding (pairwise) Type I error rates than by the timing of the addition of the new arms. 
The FWER can be estimated using \v{S}id\'{a}k's correction if the correlation between
the test statistics of pairwise comparisons is less than $0.30$.

\parttitle{Conclusions} 
The findings we present in this article can be used to design trials with pre-planned
deferred arms or to design new pairwise comparisons within an ongoing platform trial
where control of the pairwise error rate (PWER) or FWER (for a subset of pairwise comparisons)
is required. 

\end{abstract}




\end{abstractbox}
\end{fmbox}

\end{frontmatter}




\section{Introduction}

Many recent developments in clinical trials are aimed at speeding up
research by making better use of resources. Phase III clinical trials can
take several years to complete in many disease areas, requiring considerable
resources. During this time, a new promising treatment which needs to be
tested may emerge. The practical advantages of incorporating such a new
experimental arm into an existing trial protocol have been clearly described
in \cite{sydes2012} \cite{Elm2012} \cite{cohen2015} \cite{Ventz17}, not 
least because it obviates the often lengthy process of initiating a new trial and 
competing for patients. One trial using this approach is the STAMPEDE trial 
\cite{sydes2009} in men with high-risk prostate cancer. STAMPEDE is a multi-arm 
multi-stage (MAMS) platform trial that was initiated with one common control arm 
and five experimental arms assessed over four stages. Five new experimental arms 
have been added since its conception \cite{sydes2012} - see Section \ref{Example} 
for further details. This was done within the paradigm of a `platform' that has a single 
master protocol in which multiple treatments are evaluated over time. It offers flexible
features such as early stopping of accrual to treatments for
lack-of-benefit, or adding new research treatments to be tested during the course of
a trial. There might also be scenarios when at the design stage of a new trial another 
experimental arm is planned to be added after the start of the trial, i.e. a 
planned addition. An example of this scenario is the RAMPART trial in renal cancer - 
see Section \ref{Results}. In some platform designs, however, the addition of the new 
experimental arm would be intended but not specially planned at the start of the platform, 
i.e. unplanned, and is opportunistic at a later stage.

The Type I error rate is one of the key quantities in the design of any
clinical trial. Two measures of Type I error in a multi-arm trial are the
pairwise (PWER) and familywise (FWER) type I error rates. The PWER is the
probability of incorrectly rejecting the null hypothesis for the primary outcome of a particular
experimental arm at the end of the trial, regardless of other experimental
arms in the trial. The FWER is the probability of incorrectly rejecting the
null hypothesis for the primary outcome for at least one of the experimental
arms from a set of comparisons in a multi-arm trial. It gives the Type I
error rate for a set of pairwise comparisons of the experimental arms with
the control arm. In trials with multiple experimental arms the maximum
possible FWER often needs to be calculated and known - see \cite{wason2014fwer} for details. In some multi-arm trials,
this maximum value needs to be controlled at a pre-defined level. This is
called a \textit{strongly} controlled FWER as it covers all eventualities,
i.e. all possible hypotheses \cite{Bratton2016}. Dunnett \cite{dunnett1955} developed an
analytical formula to calculate the FWER in multi-arm trials when all the
pairwise comparisons of experimental arms against the control arm start and
conclude at the same time. However, it has been unclear how to calculate the FWER 
when new experimental arms are added during the course of a trial.

\begin{figure*}[t!]
\includegraphics[scale=0.68]{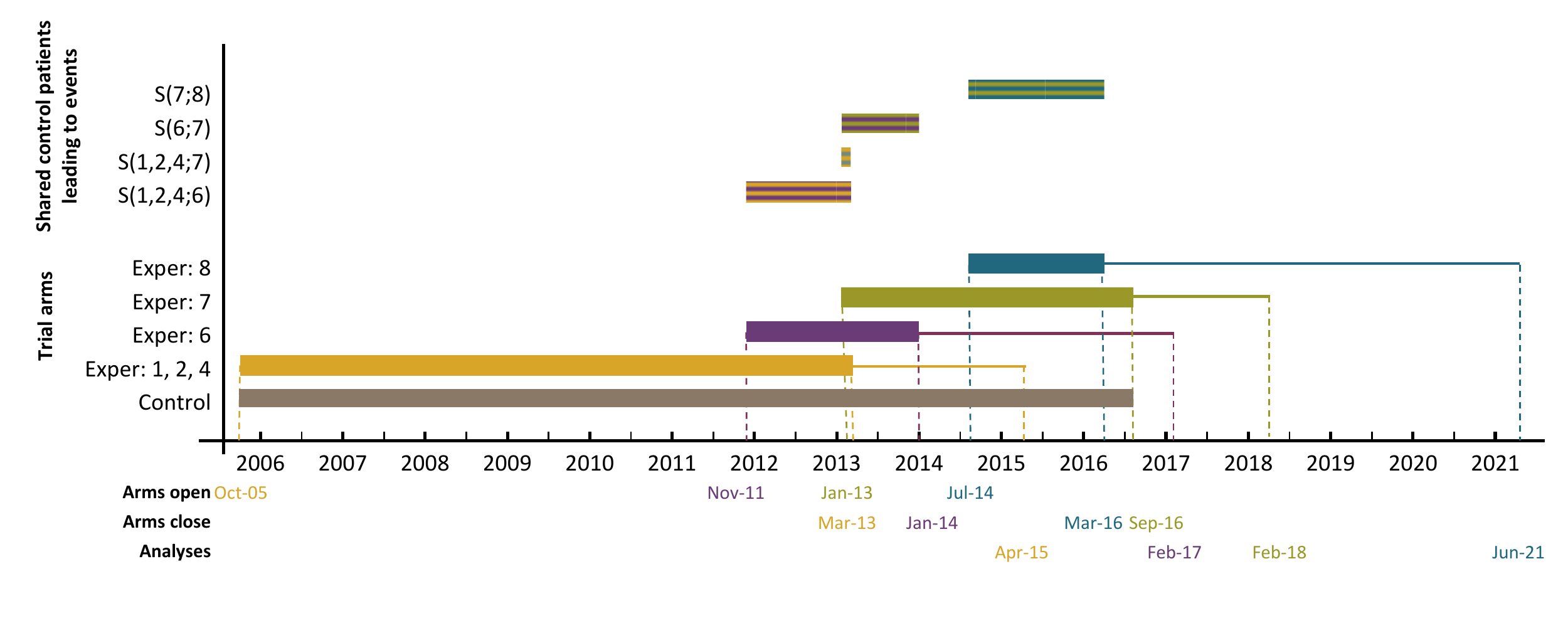} 
\onecolumn
\caption{Schematic representation of
the control and experimental arm timelines in the STAMPEDE trial. Below
section: the thick horizontal bars represent the accrual period, and the
following solid lines represent the follow-up period. Top section: the
striped bars represent the period when the recruited control arm patients overlap during this period 
between different pairwise comparisons. The colours of the stripes represent
the colours of each pairwise comparison. For example, the striped bar that is 
labelled as $S(1,2,4;6)$ represents the period when the recruited control arm patients
are shared between the original pairwise comparisons $1,2,$ and $4$ and the $6$th 
newly added comparison during this period.} 
\twocolumn
\label{fig:STAMPEDE_time}
\end{figure*}

The purpose of this article is threefold. First, we describe how the FWER, 
disjunctive (any-pair) and conjucntive (all-pairs) powers - see Section \ref{Methods} 
for their definitions - can be calculated 
when a new experimental arm is added during the course of
an existing trial with continuous, binary and time-to-event outcomes. 
Second, we describe how the FWER can be \textit{strongly} controlled at a
prespecified level for a set of pairwise comparisons in both planned (i.e.
the added arm is planned at the design stage) and unplanned (e.g. such as platform
designs) scenarios. Third, we explain how the decision to control the PWER or the FWER in a
particular design involves a subtle balancing of both practical and
statistical considerations \cite{Cook1996}. This article outlines these
issues, and provides guidance on whether to emphasise the PWER or the FWER
in different design scenarios when adding a new experimental arm. 

The structure of the article is as follows. In Section \ref{Example}, the
design of the STAMPEDE platform trial is presented. In Section %
\ref{Methods}, Methods, we explain how the FWER, disjunctive and conjunctive powers are
computed when a new experimental arm is added to
an ongoing trial. In Section \ref{Results}, Results, we present the outcome
of our simulations to verify our analytical derivation. We also show two
applications in both planned (i.e. RAMPART trial in renal cancer) and
platform design (i.e. STAMPEDE trial) settings. In Section \ref{Strategies},
based on our empirical investigations, we propose strategies that can be
applied to \textit{strongly} control the FWER when adding new experimental arms to an ongoing
platform trial in scenarios where such a control is required. Finally,
Section \ref{Discussion} is a discussion.

\section{An example: STAMPEDE trial \label{Example}}

STAMPEDE \cite{sydes2012} is a multi-arm multi-stage (MAMS) platform trial
 for men with prostate cancer at high risk of
recurrence who are starting long-term androgen deprivation therapy \cite%
{sydes2012}. In a four-stage design, five experimental arms with treatment
approaches previously shown to be safe were compared with a control arm
regimen. In this trial, the primary analysis was carried out at the end of
stage $4$, with overall survival as the primary outcome. Stages $1$ to $3$
used an intermediate outcome measure of failure-free survival. As a result,
the corresponding hypotheses at interim stages played a subsidiary role -
i.e. used for lack-of-benefit analysis  on an intermediate outcome, not for making claims of efficacy. We, 
therefore, focus on the primary hypotheses on overall survival at the final stage - as outlined below.

Recruitment to the original design began late in $2005$ and was completed
early in $2013$. The design parameters for the primary outcome at the final
stage were a (one-sided) significance level of $0.025$, power of $0.90$, and
the target hazard ratio of $0.75$ on overall survival which required $401$ control arm deaths
(i.e. events on overall survival). An allocation ratio of $A=0.5$ was used
for the original comparisons so that, over the long-term, one patient was
allocated to each experimental arm for every two patients allocated to
control. Because distinct hypotheses were being tested in each of the five
experimental arms, the emphasis in the design for STAMPEDE was on the
pairwise comparisons of each experimental arm against control, with emphasis
on the strong control of the PWER. Out of the initial five experimental
arms, only three of them continued to recruit through to their final stage.
Recruitment to the other two arms stopped at the second interim look due to lack\ of sufficient activity.

Since November $2011$, five new experimental arms have been added to the
original design. Figure \ref{fig:STAMPEDE_time} presents the timelines for
different arms, including three of those added later. [Figure \ref%
{fig:STAMPEDE_time} NEAR\ HERE]. Note that patients allocated to a new
experimental arm are only compared with patients randomised to the
control arm contemporaneously, and recruitment to the new experimental arm(s) 
continue for as long as is required. Therefore, the analysis and reporting of the new comparisons
will be later than for the original comparisons. Figure \ref{fig:STAMPEDE_time}
also shows the recruitment periods when the pairwise comparisons of the
newly-added experimental arms with the control overlap with each other as
well as with those of the original comparisons - see the top section of
Figure \ref{fig:STAMPEDE_time}.

\section{Methods\label{Methods}}

In this section, we first present the formulae for the correlation of the two
test statistics when one of the comparisons is added later in trials with continuous, 
binary and survival outcomes. We then describe how
Dunnett's test can be extended to compute the FWER, as well as conjunctive
and disjunctive powers when a new arm is added mid-course of a two arm trial.

\subsection{Type I error rates when adding a new arm}

In a two arm trial, the primary comparison is between the control group ($C$%
) and the experimental treatment ($E$). The parameter $\theta $ represents
the difference in the outcome measure between the two groups. In the notation of this article,
the control group is always identified with subscript $0$. For continuous
outcomes, $\theta $ could be the difference in the means of the two groups $%
\mu _{1}-\mu _{0}$; for binary data difference in the proportions $%
p_{1}-p_{0}$; for survival data a log hazard ratio ($\log HR$).

The efficient score statistic for $\theta $ (based on the available data and
calculated under the null hypothesis that $\theta =0$) is represented by $S$
with $V$ being the Fisher's (observed) information about $\theta $ contained
in $S$. Conditionally on the value of $V$, in large samples (which is the
underlying assumption throughout this article), $S$ follows the normal
distribution with mean $\theta V$ and variance $V$, i.e. $S\thicksim N($ $%
\theta V,V)$. In the survival case, $S$ and $V$ are the logrank test
statistic and its null variance, respectively.

In practice, the progress of a trial can be assessed in terms of
`information time' $t$ because it measures how far through the trial we are 
\cite{fallmann1994}. In the case of continuous and binary outcomes, $t$ is
defined as the total number of individuals accrued so far divided by the
total sample size. In survival outcomes, it is defined as the total number
of events occurred so far divided by the total number of events required by
the planned end of the trial \cite{fallmann1994}. In all cases $t=0$ and $%
t=1 $ correspond to the beginning and end of the trial, respectively. In
continuous, binary and survival outcome data, $S$ has independent and
normally distributed increment structure. This means that at information
times $t_{1}$ , $t_{2}$, $...$, $t_{j}$ the increments $S(t_{_{1}})$, $%
S(t_{_{2}})-S(t_{_{1}})$, $...$, $S(t_{_{j}})-S(t_{j-_{1}})$ are
independently and normally distributed.

Furthermore, the $Z$ test statistic can be expressed in terms of the
efficient score statistic $S$ and Fisher's information as $Z=S/\sqrt{V}$.
The $Z$ test statistic is (approximately) normally distributed $Z\thicksim
N( $ $\theta \sqrt{V},1)$ and has the same independent increment property as
that of $S$. For example in trials with continuous outcomes, where the aim
is to test that the outcome of $n_{1}$ individuals in experimental treatment 
$E_{1}$ is on average smaller (here smaller means better, e.g. blood
pressure) than that of $n_{0}$ individuals in control group ($C$), the null
hypothesis $H_{0}^{1}:\mu _{1}\geq \mu _{0}$ is tested against the
(one-sided) alternative hypothesis $H_{1}^{1}:\mu _{1}<\mu _{0}$. In this
case, the Type I error rate is a predefined value $\alpha _{1}=\Phi
(z_{\alpha _{1}})$ where $\Phi (.)$ is the normal probability distribution
function. Denote $Z_{1}$ the standardised test statistics for $E_{1}$ versus control. 
Under $H_{0}$, the distribution of the $Z$ test statistics is standard normal, 
$N(0,1)$. Table \ref{Outcomes} presents the test statistics for continuous,
binary, and survival outcomes with the corresponding Fisher's (observed)
information.

\begin{table*}[t] \centering%
\caption{{\it Treatment effects, test statistics, expected information, and correlation between the test statistics of pairwise 
comparisons in trials with continuous, binary, and survival outcomes, with common allocation ratio ($A$).}}%
\begin{tabular}{ccclcc}
\hline
{\small Outcome} & {\small Treatment effect} & {\small Test statistics} & 
{\small Fisher's information (}$V${\small )} & \multicolumn{2}{c}{\small %
Corr. between two comparisons} \\ \cline{5-6}
& {\small (}$\theta ${\small )} & {\small (}$Z${\small )} & 
\multicolumn{1}{c}{} & {\small Complete ovlp.} & {\small Partial ovlp.} \\ 
\multicolumn{1}{l}{} & \multicolumn{1}{l}{} & \multicolumn{1}{l}{} &  & ($%
\rho _{12}$) & ($\rho _{12}^{\ast }$) \\ \hline
\multicolumn{1}{l}{\small Continuous} & \multicolumn{1}{l}{$\theta _{c}=\mu
_{1}-\mu _{0}$} & \multicolumn{1}{l}{$Z_{1}=\theta _{c}\sqrt{V_{c}}$} & $%
V_{c}=(\frac{\sigma _{0}^{2}}{n_{0}}+\frac{\sigma _{1}^{2}}{An_{0}})^{-1}$ & 
$\frac{A}{A+1}$ & $\frac{A}{A+1}.\frac{n_{0,12}}{n_{0}}$ \\ 
\multicolumn{1}{l}{} & \multicolumn{1}{l}{$\theta _{b_{1}}=p_{1}-p_{0}$} & 
\multicolumn{1}{l}{$Z_{1}=\theta _{b_{1}}\sqrt{V_{b_{1}}}$} & $V_{b_{1}}=(%
\frac{p_{0}(1-p_{0})}{n_{0}}+\frac{p_{1}(1-p_{1})}{An_{0}})^{-1}$ & $\frac{A%
}{A+1}$ & $\frac{A}{A+1}.\frac{n_{0,12}}{n_{0}}$ \\ 
\multicolumn{1}{l}{\small Binary} & \multicolumn{1}{l}{$\theta _{b_{2}}=\log
\left\{ \frac{p_{1}(1-p_{0})}{p_{0}(1-p_{1})}\right\} $} & 
\multicolumn{1}{l}{$Z_{1}=\theta _{b_{2}}\sqrt{V_{b_{2}}}$} & $V_{b_{2}}=(%
\frac{1}{n_{0}p_{0}(1-p_{0})}+\frac{1}{An_{0}p_{1}(1-p_{1})})^{-1}$ & $\frac{%
A}{A+1}$ & $\frac{A}{A+1}.\frac{n_{0,12}}{n_{0}}$ \\ 
& \multicolumn{1}{l}{$\theta _{b_{3}}=\log \left\{ \frac{p_{1}}{p_{0}}%
\right\} $} & \multicolumn{1}{l}{$Z_{1}=\theta _{b_{3}}\sqrt{V_{b_{3}}}$} & $%
V_{b_{3}}=(\frac{1-p_{0}}{n_{0}p_{0}}+\frac{1-p_{1}}{An_{0}p_{1}})^{-1}$ & $%
\frac{A}{A+1}$ & $\frac{A}{A+1}.\frac{n_{0,12}}{n_{0}}$ \\ 
\multicolumn{1}{l}{\small Survival} & \multicolumn{1}{l}{$\theta _{s}=\log
(HR)$} & \multicolumn{1}{l}{$Z_{_{1}}=\theta _{s}\sqrt{V_{s}}$} & $%
V_{s}=(d\frac{A}{(1+A)^{2}})^{-1}$ & $\frac{A}{A+1}$ & $\frac{A}{A+1}.%
\frac{e_{0,12}}{e_{0}}$ \\ \hline
\multicolumn{6}{l}{$\rho _{12}$: correlation when there is complete
overlap between pairwise comparisons.} \\ 
\multicolumn{6}{l}{$\rho _{12}^{\ast }$: correlation when only $%
n_{0,12}$($e_{0,12}$)\ control arm observation(events) overlap between comparisons 1 and 2.} \\ 
\multicolumn{6}{l}{$n_{0,12}$($e_{0,12}$), shared observations(events) in
control arm; $n_{0}$($e_{0}$), total observations(events) in control arm; $d$%
, all events}%
\end{tabular}%
\label{Outcomes}%
\end{table*}%

If a different experimental arm $E_{2}$ is compared with the control
treatment $C$ in another independent trial, the corresponding null
hypothesis is $H_{0}^{2}:\mu _{2}\geq \mu _{0}$ with the Type I error rate
being similarly defined as $\alpha _{2}$. Magirr \textit{et al.} \cite{magirr2012} showed that the FWER is 
maximised under the global null hypothesis, $H_{0}^{G}$, that is, when the mean outcome 
in each of the experimental arms is equal to that of the control arm, $H_{0}^{G}:\theta_{1}^{0}=\theta_{2}^{0}=0$.  
In the above scenario, since the two trials are independent, the overall Type I error rate (FWER) of the two
comparisons, $k=1,2$, can be calculated using the \v{S}id\'{a}k formula \cite{Sidak67}%
\begin{eqnarray*}
FWER_{S} &=&\Pr (\text{reject at least one }H_{0}^{k}\text{ }|H_{0}^{G}) \\
&=&\Pr (\text{reject }H_{0}^{1}\text{ or }H_{0}^{2}|H_{0}^{G}) \\
&=&1-\Pr (\text{accept }H_{0}^{1}\text{ and }H_{0}^{2}|H_{0}^{G}) \\
&=&1-(1-\alpha _{1})(1-\alpha _{2}).
\end{eqnarray*}%
When $\alpha _{1}=\alpha _{2}=\alpha$, 
\begin{equation}
FWER_{S}=1-(1-\alpha )^{2}  \label{Sid_formula}
\end{equation}%
where subscript $S$ stands for \v{S}id\'{a}k. If the control arm observations are shared between the two pairwise
comparisons, one can replace the term $(1-\alpha )^{2}$ in eqn. (\ref{Sid_formula}) to 
allow for the correlation between the two test statistics $%
Z_{1}$ and $Z_{2}$, i.e. the correlation induced by the shared control
arm information. Dunnett \cite{dunnett1955} provided an analytical formula
to estimate the FWER when all the comparisons start and conclude at the same
time, i.e. when all control arm observations overlap between different
comparisons. In the above scenario, the FWER can be calculated using 
\begin{equation}
FWER_{D}=1-\Phi _{2}(z_{1-\alpha _{1}},z_{1-\alpha _{2}};\rho _{12}) \label{Dunnett_prob}
\end{equation}%
where $\Phi _{2}(.;\rho _{12})$ is the standard bivariate normal probability
distribution function and $\rho _{12}$ is the correlation between $%
Z_{1}$ and $Z_{2}$ at the final analysis. With equal allocation ratio $A=A_{1}=A_{2}$ across all experimental arms, 
 $\rho _{12}=\frac{A}{%
A+1}$ \cite{dunnett1955} - e.g. $\rho _{12}=0.5$ when $%
n_{0}=n_{1}=n_{2}$.

The formula for $\rho _{12}$ can be extended for the scenario when $%
E_{2}$ is started later than $E_{1}$ and $C$ overlaps with both of them, i.e. when only some of the control arm
observations are shared between the two comparisons. This scenario is quite
common in platform trials where new experimental arms can be added to the
previous sets of pairwise comparisons, and recruitment to the new experimental and 
control arms continues until the planned end of that comparison. To achieve this, we make use of the
asymptotic properties of the efficient score statistic and the $Z$ test
statistic. It has been shown that over time the sequence of $Z$ test
statistics approximately has an independent and normally distributed
increment structure for the estimators of the treatment effects presented in
Table \ref{Outcomes} \cite{fallmann1994} \cite{Tsiatis1981}. This means that
at information time $t^{\prime }>t$, 
\begin{equation*}
\sqrt{N(t^{\prime })}Z(t^{\prime })=\sqrt{N(t)}Z(t)+\sqrt{N(t^{\prime })-N(t)%
}Z(t^{\prime }-t)
\end{equation*}%
where $N(t^{\prime })$ and $N(t)$ are the total sample sizes at information times $%
t^{\prime }$and $t$. With equal allocation ratio to both experimental arms,
if $n_{0,12}$ control arm observations (where\ $\,0<n_{0,12}<n_{0}$) are shared
between the two comparisons, the correlation between $Z_{1}$ and $Z_{2}$ can
be calculated using eqn. (\ref{Rho_Z1Z2}) - see the online document Supplementary Material for analytical
derivations and more complex formula for the case of unequal allocation
ratio between comparisons.    
\begin{equation}
\rho _{12}^{\ast }=\frac{A}{A+1}.\frac{n_{0,12}}{n_{0}}
\label{Rho_Z1Z2}
\end{equation}%
Note that the factor $\frac{n_{0,12}}{n_{0}}$ is bounded by $\left[ 0,1\right] 
$ with the upper bound equal to $1$ when $n_{0,12}=n_{0}$ (i.e. when the two
comparisons start and finish at the same time), and the lower bound equal to 
$0$ when there is no shared observation in the control arm - in which case, $%
FWER_{D}$ converges to $FWER_{S}$. 

Our analytical derivation shows that eqn. (\ref{Rho_Z1Z2}) applies to
both continuous and binary outcome measures. However, in survival outcomes
the ratio $\frac{n_{0,12}}{n_{0}}$ should be replaced with the ratio of the
shared events in the control arm, i.e. $\frac{e_{0,12}}{e_{0}}$ - see Supplementary Material for 
analytical derivations and also the more complicated formula for unequal allocation 
ratio. Table \ref{Outcomes} shows the corresponding
formula for $\rho _{12}^{\ast }$ by the type of outcome measure.

\subsection{Power when adding a new arm}

The power of a clinical trial is the probability that under a particular
target treatment effect $\theta^{1}$, a truly effective treatment is
identified at the final analysis. In multi-arm designs, per-pair (pairwise) power 
($\omega $) \cite{royston2011} calculates this probability for a given experimental 
arm against the control. In multi-arm settings, however, there are other definitions 
of power that might be of interest - depending on the objective of the trial. In 
the above setting where there are two comparisons, define the target treatment effects under 
the alternative hypothesis for each of the comparisons as $\theta_{1}^{1}$ and $\theta_{2}^{1}$, 
respectively. Disjunctive (any-pair) power is the probability of showing a statistically 
significant effect under the targeted effects for at least one comparison: 

\begin{eqnarray*}
\Omega _{d} &=&\Pr (\text{reject at least one }H_{0}^{k}\text{ }|\theta_{1}=\theta_{1}^{1} \text{,} \theta_{2}=\theta_{2}^{1})  \\
&=&1-\Pr (\text{accept }H_{0}^{1}\text{ and }H_{0}^{2}|\theta_{1}=\theta_{1}^{1} \text{,} \theta_{2}=\theta_{2}^{1}) 
\end{eqnarray*}%
When the
two comparisons, $k=1,2$, are independent, i.e. $\rho _{12}=0$, disjunctive
power ($\Omega _{d}$) is defined as
\begin{equation}
\Omega _{d}=1-(1-\omega _{1})(1-\omega _{2}).
\end{equation}%
If $\rho _{12}\neq 0$, then $\Omega _{d}$ is calculated using 
\begin{equation}
\Omega _{d}=1-\Phi _{2}(z_{1-\omega _{1}},z_{1-\omega _{2}};\rho_{12})  \label{Disj_Power}
\end{equation}

Conjunctive (all-pairs) power is the probability of showing a statistically
significant effect under the targeted effects for all comparison pairs. When
the two tests are independent, conjunctive power ($\Omega _{c}$) is

\begin{eqnarray*}
\Omega _{c} &=&\Pr (\text{reject all }H_{0}^{k}\text{ }|\theta_{1}=\theta_{1}^{1} \text{,} \theta_{2}=\theta_{2}^{1}) \\
&=&\Pr (\text{reject }H_{0}^{1}\text{ and }H_{0}^{2}|\theta_{1}=\theta_{1}^{1} \text{,} \theta_{2}=\theta_{2}^{1}) \\
&=&\omega _{1}.\omega _{2}
\end{eqnarray*}%
Given the correlation $\rho_{12}$, then $\Omega _{c}$ is calculated
using 
\begin{equation}
\Omega _{c}=\Phi _{2}(z_{\omega _{1}},z_{\omega _{2}};\rho_{12})
\label{Conj_Power}
\end{equation}%
If a new experimental arm is added later on, the corresponding formula for $\rho
_{12}^{\ast }$ in Table \ref{Outcomes} can be used to calculate
both disjunctive ($\Omega _{d}$) and conjunctive ($\Omega _{c}$) powers in
this scenario.

\section{Results\label{Results}}

In this section, we first show the results of our simulations to explore the
validity of eqn. (\ref{Rho_Z1Z2}) to estimate $\rho_{12}^{\ast }$,
and to study the impact of the timing of the addition of a new experimental
arm on the FWER and different types of power. Because of the censoring, survival outcomes 
are generally considered the most complex type of outcomes listed in Table \ref{Outcomes}. 
We conduct our simulations in this setting. Then, we estimate the
correlation structure between the test statistics of different comparisons
in the STAMPEDE trial, including the first three of the added arms to the original set of
comparisons. Finally, to illustrate the design implications in planned
scenarios, we show an application in the design of the RAMPART trial.

\subsection{Simulation design}

In our simulations, we considered a hypothetical three-arm trial with one
control, $C$, and two experimental arms ($E_{1}$ and $E_{2}$). We applied
similar design parameters to those in\textit{\ }\cite{royston2011} - see
Section 2.7.1 - taking median survival for the time-to-event outcome of $1$
yr (hazard $\lambda _{1}=0.693$ in control arm). We generated individual
time-to-event patient data from an exponential distribution and estimated
the correlation between the test statistics of the two pairwise comparisons $%
Z_{1}$ and $Z_{2}$ when $E_{2}$ was initiated at different time points after
the start of the experimental arm $E_{1}$ and control. Accrual rates were
assumed to be uniform throughout (across the platform) and set to $500$
patients per unit time for both comparisons.

As in the STAMPEDE trial, the comparison set of patients for the deferred
experimental arm $E_{2}$ are the contemporaneously-recruited control arm $C$
individuals. This means that in our simulations recruitment to the control
arm continued until conclusion of that required for the $E_{2}$ comparison.
As for the final stage of STAMPEDE, the design significance level and power
were chosen as $\alpha _{i}=0.025$ and $\omega _{i}=0.9$, $i=1,2$ in all
scenarios. The target hazard ratio under the alternative hypothesis in both
pairwise comparisons were $0.75$. To investigate the FWER under different
allocation ratios, we carried out our simulations under three allocation
ratios of $A=\left\{ 0.5,1,2\right\} $. Table \ref{A} [Table \ref{A} NEAR\
HERE] presents details of the design parameters, including trial timelines,
in each pairwise comparison for different values of $A$. Calculations for
Table \ref{A} were done in Stata using the \texttt{nstage} program \cite%
{bratton2015}. In simulations, $50,000$ replications were generated in each
scenario.

Finally, the main aim of our simulation study is to explore the impact of the timing 
of adding a new experimental arm on the correlation structure and the value of the FWER. 
For this reason, only one original comparison was included in our simulations. 
In Section \ref{Strategies} and Discussion, we discuss how the FWER can be strongly 
controlled, and address other relevant design issues, when more pairwise comparisons 
start at the begining.

\begin{table}[tbp] \centering%
\caption{{\it Three different trial designs for each pairwise comparison of
experimental arm versus control in a 3-arm trial.}}

\begin{tabular}{ccccc}
\hline
Scenario & $A$ & $e_{0}$ & $n_{0}$ & Overall trial period \\ \hline
$1$ & $0.5$ & $401$ & $788$ & $2.33$ \\ 
$2$ & $1$ & $264$ & $545$ & $2.18$ \\ 
$3$ & $2$ & $196$ & $389$ & $2.36$ \\ \hline
\multicolumn{5}{l}{Key: $A$, allocation ratio; $e_{0}$, total control arm events required;} \\ 
\multicolumn{5}{l}{$n_{0}$, number of patients accrued to control arm by the end
of trial;} \\ 
\multicolumn{5}{l}{Overall trial period, duration (in time units) up to the
final analysis.}%
\end{tabular}%
\label{A}%
\end{table}%

\begin{table*}\centering%

\caption{{\it Estimates of the correlation between the test-statistics of the two pairwise comparisons, $Z_{1}$ and $Z_{2}$, by the timing of the addition of experimental arm $E_{2}$. 
The values for $\rho _{12}^{\ast }$ are calculated from eqn. (3). The estimates $\widehat{\rho }_{12}^{\ast }$
are obtained from simulating individual patient data.
The number of trial-level replicates is 50,000 in all experimental conditions.}}%
\begin{tabular}{cccrccccccrccrr}
\hline
& \multicolumn{4}{c}{$Allocation\ Ratio\ =\ 0.5$} &  & 
\multicolumn{4}{c}{$Allocation\ Ratio\ =1$} &  & 
\multicolumn{4}{c}{$Allocation\ Ratio\ =\ 2$} \\ 
\cline{2-5}\cline{7-10}\cline{12-15}
{\small Time $E_{2}$} & {\small Shared ctrl.}   &  &  &  &  & {\small Shared ctrl.}  &  &  &  &  & {\small Shared ctrl.}  &  &  &  \\ 
{\small started} & {\small arm events} & $\rho _{12}^{\ast }$ & $\widehat{\rho }%
_{12}^{\ast }$ & {\small FWER} &  & {\small arm events} & $\rho _{12}^{\ast }$
& $\widehat{\rho }_{12}^{\ast }$ & {\small FWER} &  & {\small arm events} & $\rho
_{12}^{\ast }$ & \multicolumn{1}{c}{$\widehat{\rho }%
_{12}^{\ast }$} & {\small FWER} \\ \cline{2-5}\cline{7-10}\cline{12-15}
&  &  & \multicolumn{1}{c}{} &  &  &  &  &  & \multicolumn{1}{r}{} &  & 
\multicolumn{1}{r}{} & \multicolumn{1}{r}{} & \multicolumn{1}{c}{} &  \\ 
$0.0$ & $401$ & $0.33$ & \multicolumn{1}{c}{$0.33$} & $0.047$ &  & $264$ & $%
0.50$ & $0.50$ & $0.045$ & \multicolumn{1}{c}{} & $196$ & $0.66$ & 
\multicolumn{1}{c}{$0.66$} & \multicolumn{1}{c}{$0.043$} \\ 
$0.2$ & $349$ & $0.29$ & \multicolumn{1}{c}{$0.29$} & $0.048$ &  & $226$ & $%
0.43$ & $0.43$ & $0.046$ & \multicolumn{1}{c}{} & $170$ & $0.57$ & 
\multicolumn{1}{c}{$0.57$} & \multicolumn{1}{c}{$0.044$} \\ 
$0.4$ & $298$ & $0.25$ & \multicolumn{1}{c}{$0.25$} & $0.048$ &  & $190$ & $%
0.36$ & $0.36$ & $0.047$ & \multicolumn{1}{c}{} & $144$ & $0.49$ & 
\multicolumn{1}{c}{$0.49$} & \multicolumn{1}{c}{$0.045$} \\ 
$0.6$ & $249$ & $0.20$ & \multicolumn{1}{c}{$0.20$} & $0.048$ &  & $155$ & $%
0.29$ & $0.29$ & $0.048$ & \multicolumn{1}{c}{} & $121$ & $0.41$ & 
\multicolumn{1}{c}{$0.41$} & \multicolumn{1}{c}{$0.046$} \\ 
$0.8$ & $204$ & $0.17$ & \multicolumn{1}{c}{$0.18$} & $0.049$ &  & $123$ & $%
0.23$ & $0.23$ & $0.048$ & \multicolumn{1}{c}{} & $98$ & $0.33$ & 
\multicolumn{1}{c}{$0.33$} & \multicolumn{1}{c}{$0.047$} \\ 
$1.0$ & $161$ & $0.13$ & \multicolumn{1}{c}{$0.14$} & $0.049$ &  & $94$ & $%
0.18$ & $0.18$ & $0.049$ & \multicolumn{1}{c}{} & $77$ & $0.26$ & 
\multicolumn{1}{c}{$0.26$} & \multicolumn{1}{c}{$0.048$} \\ 
$1.2$ & $122$ & $0.10$ & \multicolumn{1}{c}{$0.10$} & $0.049$ &  & $68$ & $%
0.13$ & $0.13$ & $0.049$ & \multicolumn{1}{c}{} & $57$ & $0.19$ & 
\multicolumn{1}{c}{$0.19$} & \multicolumn{1}{c}{$0.049$} \\ 
$1.4$ & $88$ & $0.07$ & \multicolumn{1}{c}{$0.07$} & $0.049$ &  & $45$ & $%
0.09$ & $0.09$ & $0.049$ & \multicolumn{1}{c}{} & $41$ & $0.14$ & 
\multicolumn{1}{c}{$0.14$} & \multicolumn{1}{c}{$0.049$} \\ 
$1.6$ & $57$ & $0.05$ & \multicolumn{1}{c}{$0.05$} & $0.049$ &  & $26$ & $%
0.05$ & $0.05$ & $0.049$ & \multicolumn{1}{c}{} & $26$ & $0.09$ & 
\multicolumn{1}{c}{$0.09$} & \multicolumn{1}{c}{$0.049$} \\ 
$1.8$ & $33$ & $0.03$ & \multicolumn{1}{c}{$0.03$} & $0.049$ &  & $12$ & $%
0.02$ & $0.02$ & $0.049$ & \multicolumn{1}{c}{} & $15$ & $0.05$ & 
\multicolumn{1}{c}{$0.05$} & \multicolumn{1}{c}{$0.049$} \\ 
$2.0$ & $14$ & $0.01$ & \multicolumn{1}{c}{$0.02$} & $0.050$ &  & $3$ & $0.00
$ & $0.00$ & $0.050$ & \multicolumn{1}{c}{} & $6$ & $0.02$ & 
\multicolumn{1}{c}{$0.02$} & \multicolumn{1}{c}{$0.049$} \\ \hline
\end{tabular}%
\label{simresult1}

\end{table*}%

\begin{table*}[t] \centering%
\caption{{\it Disjunctive ($\Omega _{d}$) and conjunctive ($\Omega _{c}$) powers by the timing of the addition of the 
second arm and the correlation between the test statistics of the two pairwise comparisons.}}%
\begin{tabular}{cccrccccrccr}
\hline
& \multicolumn{3}{c}{$Allocation\ Ratio\ =\ 0.5$} &  & \multicolumn{3}{c}{$%
Allocation\ Ratio\ =1$} &  & \multicolumn{3}{c}{$Allocation\ Ratio\ =\ 2$}
\\ \cline{2-4}\cline{6-8}\cline{10-12}
Time $E_{2}$ &  &  &  &  &  &  &  &  &  &  &  \\ 
started & $\rho _{12}^{\ast }$ & $\Omega _{d}$ & \multicolumn{1}{c}{$%
\Omega _{c}$} &  & $\rho _{12}^{\ast }$ & $\Omega _{d}$ & $\Omega
_{c}$ & \multicolumn{1}{c}{} & $\rho _{12}^{\ast }$ & $\Omega _{d}$
& \multicolumn{1}{c}{$\Omega _{c}$} \\ 
\cline{2-4}\cline{6-8}\cline{10-12}
&  &  & \multicolumn{1}{c}{} &  &  &  &  &  & \multicolumn{1}{r}{} & 
\multicolumn{1}{r}{} & \multicolumn{1}{c}{} \\ 
$0.0$ & $0.33$ & $0.977$ & \multicolumn{1}{c}{$0.823$} &  & $0.50$ & $0.968$
& $0.833$ & \multicolumn{1}{c}{} & $0.66$ & $0.956$ & $0.844$ \\ 
$0.2$ & $0.29$ & $0.979$ & \multicolumn{1}{c}{$0.821$} &  & $0.43$ & $0.972$
& $0.828$ & \multicolumn{1}{c}{} & $0.57$ & $0.963$ & $0.837$ \\ 
$0.4$ & $0.25$ & $0.980$ & \multicolumn{1}{c}{$0.819$} &  & $0.36$ & $0.975$
& $0.825$ & \multicolumn{1}{c}{} & $0.49$ & $0.968$ & $0.832$ \\ 
$0.6$ & $0.20$ & $0.983$ & \multicolumn{1}{c}{$0.817$} &  & $0.29$ & $0.979$
& $0.821$ & \multicolumn{1}{c}{} & $0.41$ & $0.972$ & $0.827$ \\ 
$0.8$ & $0.17$ & $0.984$ & \multicolumn{1}{c}{$0.816$} &  & $0.23$ & $0.982$
& $0.819$ & \multicolumn{1}{c}{} & $0.33$ & $0.977$ & $0.823$ \\ 
$1.0$ & $0.13$ & $0.986$ & \multicolumn{1}{c}{$0.815$} &  & $0.18$ & $0.984$
& $0.817$ & \multicolumn{1}{c}{} & $0.26$ & $0.980$ & $0.820$ \\ 
$1.2$ & $0.10$ & $0.987$ & \multicolumn{1}{c}{$0.813$} &  & $0.13$ & $0.986$
& $0.815$ & \multicolumn{1}{c}{} & $0.19$ & $0.983$ & $0.817$ \\ 
$1.4$ & $0.07$ & $0.988$ & \multicolumn{1}{c}{$0.812$} &  & $0.09$ & $0.987$
& $0.813$ & \multicolumn{1}{c}{} & $0.14$ & $0.985$ & $0.815$ \\ 
$1.6$ & $0.05$ & $0.988$ & \multicolumn{1}{c}{$0.812$} &  & $0.05$ & $0.988$
& $0.812$ & \multicolumn{1}{c}{} & $0.09$ & $0.987$ & $0.813$ \\ 
$1.8$ & $0.03$ & $0.989$ & \multicolumn{1}{c}{$0.811$} &  & $0.02$ & $0.989$
& $0.810$ & \multicolumn{1}{c}{} & $0.05$ & $0.988$ & $0.812$ \\ 
$2.0$ & $0.01$ & $0.990$ & \multicolumn{1}{c}{$0.810$} &  & $0.00$ & $0.990$
& $0.810$ & \multicolumn{1}{c}{} & $0.02$ & $0.989$ & $0.810$ \\ \hline
\end{tabular}%
\label{Power_est}%
\end{table*}%

\subsection{Simulation results}

The results are summarised in Table \ref{simresult1}. The table shows the
estimates of the correlation between the test-statistics of the two pairwise
comparisons as computed from the corresponding equation for $\rho _{12}^{\ast }$ in
Table \ref{Outcomes}, by the timing of when the second experimental arm $E_{2}$ 
was added. We estimated the number of shared control arm
events, $e_{_{0,12}}$, via simulation. For each scenario, we also simulated
individual patient data under the null hypothesis and computed both test
statistics, which were then used to estimate the correlation between them.
The results are also included in Table \ref{simresult1}, i.e. $\widehat{\rho }_{12}^{\ast }$. 
The results indicate that the estimated values of $\widehat{\rho }_{12}^{\ast }$ accord well with the corresponding
values obtained from the formula in all experimental conditions - they only
differed in the third decimal place.

The results for each allocation ratio indicate that when $E_{2}$ starts
later than $E_{1}$ and $C$, the estimates of the FWER and $\widehat{\rho }_{12}^{\ast }$ are 
driven by the shared control arm events
between the two pairwise comparisons - see Table \ref{simresult1}. The
higher the number of the shared control arm events, the lower the value of 
the FWER is because $\rho _{12}^{\ast }$ is higher. The FWER reaches 
its maximum when there is no shared information between the two pairwise 
comparisons at which point eqn. (\ref{Sid_formula}) can be used to calculate the FWER. 
This is when the two pairwise comparisons are effectively two completely independent trials 
in one protocol. In this case, Bonferroni correction can also provide a good 
approximation, i.e. $\alpha _{1}+\alpha_{2}=0.05$.The results indicate that, 
even for a correlation of as high as $0.30$, the Bonferroni correction provides
a good approximation. This correlation threshold corresponds to an overlap
(in terms of `information time') of about 60\% between the newly added
comparison and that of the existing one for an equal allocation ratio ($A=1$%
). If more individuals are allocated to the control arm (i.e. $A<1$), the
amount of overlap has to be even higher to achieve this correlation
threshold, e.g. about $87\%$ when $A<1$ in our simulations for either
experimental arm.

Furthermore, it is evident from our simulations that when more individuals
are allocated to the control arm (i.e. $A<1$), the timing of adding a new experimental 
arm has very little impact on the value of the FWER.
For an equal allocation ratio ($A=1$) the impact on the FWER is modest;
whereas for the uncommon scenario of allocation ratio $A>1$, the impact is
moderate. Therefore, in many multi-arm trials, where often more individuals are
allocated to the control arm than each experimental arm, the timing of the
addition of a new experimental arm is unlikely to be a major issue.

Finally, Table \ref{Power_est} presents the disjunctive and conjunctive powers 
in each scenario. The results indicate that the timing of the addition of the new 
arm has more impact on both types of powers. Nonetheless, the impact is still 
relatively low - particularly, when the allocation ratio is less than one. 
However, the degree of overlap affects the two types of powers in 
opposite directions. While conjunctive power decreases with smaller overlap, disjunctive 
power increases in such scenario.

\subsection{FWER of STAMPEDE when new arms were added}

In this section, we calculate the correlation between the test statistics of
different pairwise comparisons in STAMPEDE when new arms are added. The
newly added therapies look to address different research questions than
those of the original comparisons. When the first new arm was added,
STAMPEDE had only $3$ experimental arms open to accrual because arms $E_{3}$
and $E_{5}$ were stopped at their second interim look. The new experimental
arms $E_{6}$, $E_{7}$, and $E_{8}$ were added in November $2011$, January $%
2013$, and July $2014$, respectively. The final stage design parameters of
the three added comparisons were similar to those of the original
comparisons (i.e. final stage sig. level and power of $\alpha _{i}=0.025$
and $\omega _{i}=0.90$), except that their allocation ratio was set as $%
A_{6}=A_{7}=A_{8}=1$. Some of the control arm patients recruited from the
start of $E_{6}$ and $E_{7}$ are shared between the original family and the
new comparisons - see the top section of Figure \ref{fig:STAMPEDE_time}. In
all three added comparisons the final analysis takes place when $267$
primary outcome\ events are observed in the contemporaneously-randomised
control arm patients.

To calculate the correlation between different test statistics, we needed to 
estimate (or predict) the shared control arm events of the corresponding 
pairwise comparisons - Supplementary Material explains this in detail. Only $7$ common control arm 
primary outcome events were expected to be shared between $E_{7}$ and
the original family of pairwise comparisons at their respective primary analysis (see Table 1 in Supplementary Material). For the $E_{6}$ comparison, 
the (expected) number of common control arm primary outcome events is $77$, but it is still a small fraction of the
total events required at its main analysis. As a result, the correlations
between the corresponding test statistics are quite low in both cases, i.e. $\widehat{\rho }_{k6}^{\ast }=0.12$ 
and $\widehat{\rho }_{k7}^{\ast }=0.01$, $k=1,2,4$.\ Between $%
E_{6}$ and $E_{7}$ comparisons, the number of shared primary events were expected
to be higher ($e_{0,12}=92$) which will result in a slightly higher
correlation. But, even in this case the correlation is well below $0.30$.
Not only do the first three added comparisons pose distinct research
questions, the correlation between the test statistics of the
corresponding pairwise comparisons are very low. If the strong control of the 
FWER was required for the three added arms, the simple Bonferroni correction could 
have been used to approximate Dunnett's correction since both the correlation and 
the amount of overlap between the three comparisons are very low.

\begin{figure*}[t!]

\includegraphics[scale=0.6]{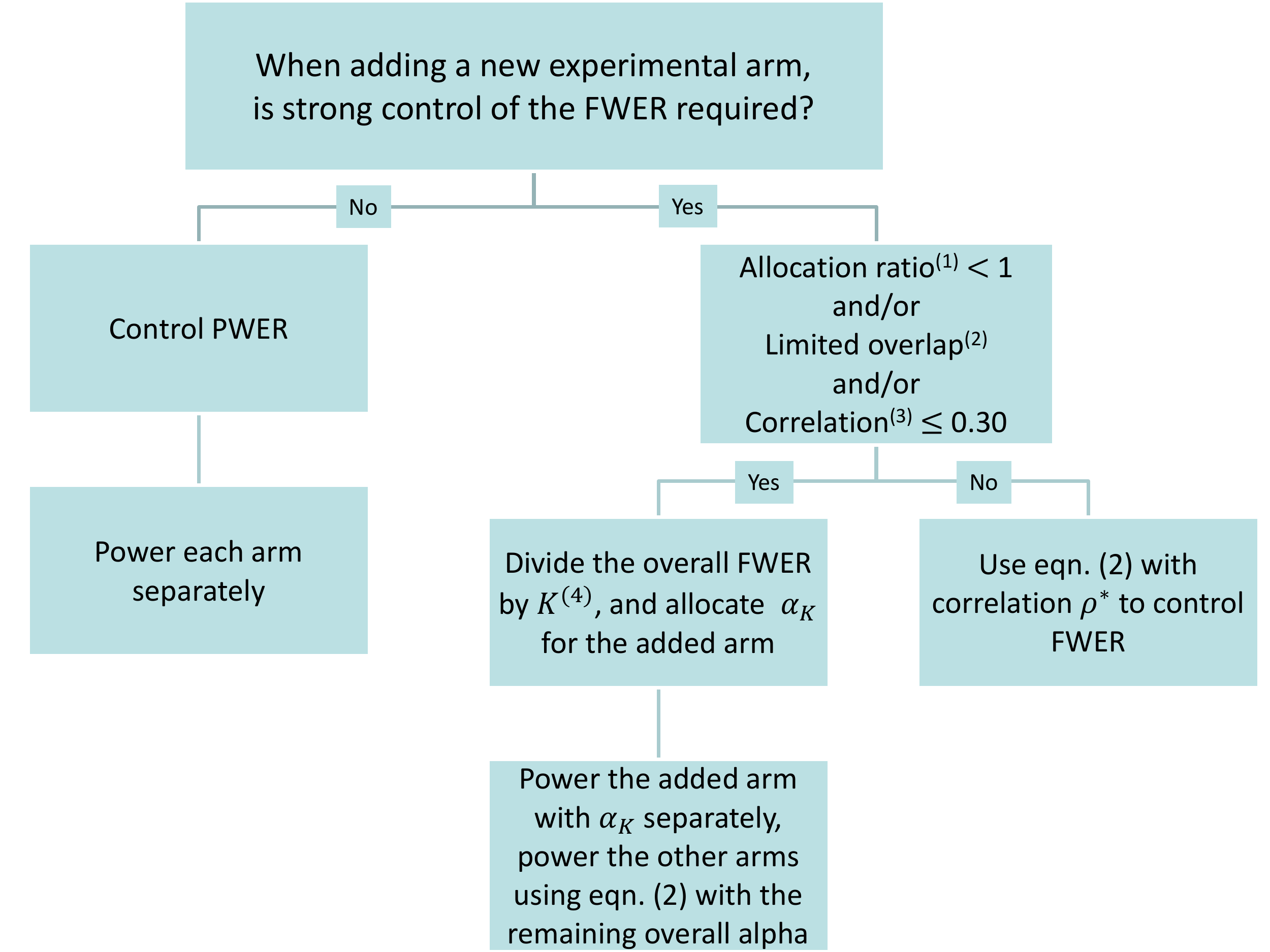}
\onecolumn
\caption{Strategies to control Type I error rate when adding new experimental arms. 
\newline Key: 1) allocation ratio for either of the new or ongoing comparisons; 
2) e.g. $<60\%$ of information time when $A=1$; 3) correlation between the test statistics of pairwise comparisons; 4) $K$ is the total number of pairwise comparisons, including the added arms.}
\twocolumn
 \label{fig:Chart}
\end{figure*}

\subsection{Design application: RAMPART trial}

Renal Adjuvant MultiPle Arm Randomised Trial (RAMPART) is an international
phase III trial of adjuvant therapy in patients with resected primary
renal cell carcinoma (RCC) at high or intermediate risk of relapse. In this multi-arm multi-stage (MAMS) 
trial, the control arm ($C$), i.e. active monitoring, and first two experimental
arms ($E_{1}$ and $E_{2}$) are due to start recruitment at the same time, with another experimental
treatment ($E_{3}$) - which is in early phase development - expected to be added at
least two years after the start of the first three arms. The deferred
experimental arm, $E_{3}$, will share some of the control arm patients with
the other two comparisons and only be compared against those recruited
contemporaneously to the control arm over the same period. No head-to-head 
comparison of the experimental arms planned, and all the stopping boundaries are prespecified. The trial design has
passed both scientific and regulatory reviews, obtained approval from both the 
EMA and FDA, and started in mid-2018. Upon reviews of the design, it 
was deemed necessary to control the FWER at $2.5\%$ 
(one-sided) in this trial, whether or not the deferred arm is added. Table 2 in Supplementary Material presents the 
design details for RAMPART - for full details of the design and trial protocol, please see 
https://www.rampart-trial.org/.

We carried out simulations to investigate the impact of the timing of adding 
the third experimental arm on the FWER. This was done at 2 years, 3
years and 4 years into the two original comparisons. The simulation results
confirmed our findings that the timing of the addition of $E_{3}$ has no
practical impact on the value of FWER. Therefore, the overall (one-sided) Type I error
rate was proportionally split between the two comparisons that start at the
same time and the deferred comparison, i.e. $E_{3}$ vs. $C$, using the Dunnett correction. 
Note that in the two comparisons that start at the same time there is a large proportion
of shared control arm information. To make use of the induced correlation
between the test statistics of these comparisons, simulations were used to
approximate Dunnett probability in this case. Simulations showed that the
final stage significance level of $0.0097$ in all pairwise comparisons
controls the overall FWER at $2.5\%$ when the deferred arm is added later
on. Our simulations also showed that the final stage significance level of
the two original pairwise comparisons can be increased to $0.015$ if $E_{3}$
is not added to buy back the unspent Type I error of the third pairwise
comparison. This will decrease the required sample size in these two
comparisons - see the last two columns of Table 2 in Supplementary Material - and will
bring forward the (expected) timing of the final analysis in both $E_{1}$ vs. $C$
comparison ($\sim 10$ months) and in $E_{2}$ vs. $C$ comparison ($\sim 4$
months).

\section{Strong control of FWER when required \label{Strategies}}

Opinions differ as to whether the FWER needs to be strongly controlled in
all multi-arm trials \cite{Cook1996} \cite{wason2014fwer} \cite{Dena2016} 
\cite{Proschan2000} \cite{Rothman1990} \cite{freidlin2008}. In our view
there are cases such as examining different doses of the same drug where the
control of the FWER might be necessary to avoid offering a particular
therapy an unfair advantage of showing a beneficial effect. However, in many 
multi-arm trials where the research treatments in the existing and added 
comparisons are quite different from each other, we believe that the greater 
focus should be on controlling each pairwise error rate \cite{OBrien83} \cite{Cook1996}. 
To support this view consider the following: if two distinct experimental 
treatments are compared to a current standard in independent trials, it is accepted that there is no
requirement for multiple testing adjustment \cite{Dena2016}. Therefore, it
seems fallacious to impose an unfair penalty if these two hypotheses are
instead assessed within the same protocol where both hypotheses are powered
separately and appropriately. This is seen most clearly when the data remain
entirely independent, for example, when these are non-overlapping with effectively 
separate control groups.

Our results indicate that the timing of adding a new experimental arm to an
ongoing multi-arm trial - where the allocation ratio is often one or
less, i.e. more patients are recruited to the control arm - is almost
irrelevant in terms of changing the value of the FWER. Even in cases where 
there is an overlap (in terms of `information time') of $60\%$ the impact on 
the increase of FWER can be negligible. The practical implication of this finding is that in cases 
where strong control of the FWER is required, one can simply divide the overall FWER by 
the total number of pairwise comparisons $K$, including the added arms, and take the worse case 
scenario of complete independence and design the deferred arm with $\alpha_{K}$ as an independent 
trial. In this case, they can be considered as separate trials and the new hypothesis 
can be powered separately and appropriately. If the FWER for the protocol as a
whole is required to be controlled at a certain level, as in the RAMPART
trial, then the overall Type I error can be split accordingly between the
original and added comparisons. This seems to be a practical strategy to
control the FWER because in most cases the exact timing of the availability
of a new experimental therapy may not be determined in advance. If the new
experimental arm is not actually added, the final stage significant level of
the original comparisons can be relaxed to achieve the target FWER. 
There might be situations where more experimental therapies are available
later than planned at the design stage. In this case, one way to control the
FWER for the new set of pairwise comparisons is to reduce the final
significance level for the existing comparisons. But, this would increase
the (effective) sample size of the existing comparisons and thus the overlap
between the new and existing comparisons would increase - which in turn
would affect FWER. In this case, a recursive procedure would be required to
achieve the desired level for the FWER.

Finally, we emphasise that the decision to control the PWER or the FWER (for
a set of pairwise comparisons) depends on the type of research questions
being posed and whether they are related in some way, e.g. testing different
doses or duration of the same therapy in which case the control of the FWER
may be required. These are mainly practical considerations and should be
determined on a case-by-case basis in the light of the rationale for the
hypothesis being tested and the aims of the protocol for the trial. Once a decision 
has been made to strongly control (or not) the FWER, Figure \ref{fig:Chart} 
summarises our guidelines on how to power the added comparison to 
guarantee strong control of the FWER. We believe this is a
logical and coherent way to assess the control of Type I error in most
scenarios.

\section{Discussion\label{Discussion}}

It is practically advantageous to add new experimental arms to an existing
trial since it not only prevents the often lengthy process of initiating a
new trial but also it helps to avoid competing trials being conducted \cite%
{sydes2012} \cite{Elm2012}. It also speeds up the evaluation of newly
emerging therapies, and can reduce costs and numbers of patients required 
\cite{cohen2015} \cite{Ventz17}. In this article, we studied the familywise
type I error rate and power when new experimental arms are added to an
ongoing trial.

Our results show that, under the design conditions, the correlation
between the test statistics of pairwise comparisons is affected by the
allocation ratio and the number of common control arm shared observations in
continuous and binary outcomes and primary outcome events in trials with
survival outcomes. The correlation decreases if more individuals are
proportionately allocated to the control arm. This correlation increases as
the proportion of shared control arm information increases, and it reaches
its maximum when the number of observations (in continuous and binary
outcomes) or events (in survival outcomes) is the same in both pairwise
comparisons. Our results also showed that the correlation between the
pairwise test statistics and the FWER are inversely related. The higher the
correlation, the lower the FWER is.

We reiterate that in a platform protocol the emphasis of the design should be
on the control of the PWER if distinct research questions are posed in each
pairwise comparison, particularly when there is little or no overlap between
the comparisons. To support this we would argue that the scientific community at large is 
increasingly judging the effects of treatments using meta-analysis rather than 
focusing on specific individual trial results \cite{Meta2015}. For this purpose, the readers and reviewers 
are not concerned about the value of Type I error for each trial or a set of such trials. 

Another relevant question in a multi-arm platform protocol is what constitutes a 
\textit{family} of pairwise comparisons. The difficulty in specifying a 
\textit{family} arises mainly due to the dynamic nature of a platform trial,
i.e. stopping of accrual to experimental treatments for lack-of-benefit,
and/or adding new treatments to be tested during the course of the trial.
The definition of a \textit{family} in this context involves a subtle
balancing of both practical and statistical considerations. The practical
and non-statistical considerations can be more complex in nature, hence the
need for (case-by-case) assessment. Moreover, therapies that emerge
over time are more likely to be distinct rather than related, for example,
different drugs entirely rather than doses of the same therapy. For this reason, each hypothesis is
more likely to inform a different claim of effectiveness of previously
tested agents. An example is the STAMPEDE platform trial where distinct
hypotheses were being tested in each of the new experimental arms, and these
do not contribute towards the same claim of effectiveness for an individual drug. In this case, the
chance of a false positive outcome for either claim of effectiveness is not
increased by the presence of the other hypothesis.

Although we have focused on single stage designs, our approach can easily be
extended to the multi-stage setting where the stopping boundaries are prespecified. As we have shown in RAMPART,
if there are interim stages in each pairwise comparison, the correlation
between the test statistics of different pairwise comparisons at interim
stages also contribute to the overall correlation structure. Similar
correlations formula to those presented in Table \ref{Outcomes} can be
analytically driven, see Supplementary Material, to calculate the interim stages
correlation structure. Our experience has shown that even in this case the
correlation between the final stage test statistics principally drives the FWER. Our
empirical investigation has indicated that even large changes in the
correlation between the interim stage test statistics have minimal impact on
the estimates of the FWER. However, if researchers wish to have the
flexibility of non-binding stopping guidelines, then the correlation
structure can be estimated in the same manner as discussed in this article by considering 
the correlation between the final stage test statistics only. 

Furthermore, in our simulations one experimental arm started with control at the begining of the trial 
since the aim was to investigate the impact of the timing of adding new experimental arms 
on the correlation structure and the value of the FWER. In many scenarios such as RAMPART 
and STAMPEDE, more than one experimental arms start at the same time in which case 
there will be substantial overlap in information between the pairwise comparison of 
these arms to control. If strong control of the FWER is required in this case, Dunnett's 
correction (i.e. eqn. (\ref{Dunnett_prob})) should be used to calculate the proportion of the 
Type I error rate that is allocated to each of these comparisons as we have done in the case of RAMPART.  

Moreover, in some designs such as RAMPART it is required to control the FWER
at a pre-specified level. In general, any unplanned adaptation would affect
the FWER of a trial. This includes the unplanned addition of a new
experimental arm. It will be possible to (strongly) control the FWER if the
addition of new pairwise comparisons is planned at the design stage of a
MAMS trial as we have shown in the RAMPART example. In this case, the
introduction of a new hypothesis will be completely independent of the
results of the existing treatments. In platform protocols in general, it becomes infeasible to
control the FWER for all pairwise comparisons as new experimental treatments
are added to the existing sets of pairwise comparisons.

In this article, we have investigated one statistical aspect of adding new
experimental arms to a platform trial. The operational and trial conduct
aspects also require careful consideration, some of which have already been
addressed in \cite{sydes2012}. Sydes \textit{et al.} \cite{sydes2012} put
forward a number of useful criteria that can be thought about when
considering the rationale for adding any new experimental arm. Both
statistical and conduct aspects require careful examination to efficiently
determine whether and when new experimental arms can be added to an existing
platform trial.

\subsection{Conclusions}

The familywise type I error rate is mainly driven by the number of pairwise comparisons 
and the corresponding pairwise Type I errors rates. The timing of adding a new experimental arm 
to an existing platform protocol can have minimal, if any, impact on the FWER. The simple Bonferroni or \v{S}id\'{a}k correction 
can be used to approximate Dunnett's correction in eqn. (\ref{Dunnett_prob}) if there is not a 
substantial overlap between the new comparison and those of the existing ones, or when the 
correlation between the test statistics of the new comparison and those of the existing
comparisons is small, less than, say, $0.30$.





\bibliographystyle{vancouver} 
\bibliography{AddArmII}      




\end{document}